\documentclass[11pt,prd,nofootinbib,reprint,superscriptaddress,longbibliography,colorlinks=true,citecolor=NavyBlue]{revtex4-1}
\usepackage{amsmath, amssymb, amsthm, graphicx, epsfig, fancyhdr,epsfig,multirow}
\usepackage[dvipsnames]{xcolor}
\usepackage{soul}
\usepackage{tikz-feynman}
\tikzfeynmanset{compat=1.1.0}

\usepackage{tikzsymbols}
\usepackage{natbib}
\usepackage{xcolor}
\usepackage{hyperref}

\makeatletter
\def\alt{\mathrel{\mathpalette\gl@align<}}
\def\agt{\mathrel{\mathpalette\gl@align>}}
\def\gl@align#1#2{\lower.6ex\vbox{\baselineskip\z@skip\lineskip\z@
\ialign{$\m@th#1\hfil##\hfil$\crcr#2\crcr\sim\crcr}}} \makeatother

\def\beq{\begin{equation}}
\def\eeq{\end{equation}}
\def\bea{\begin{eqnarray}}
\def\eea{\end{eqnarray}}

\pdfoutput=1

\begin{document}

\pagestyle{plain}

\title{Future Constraints on Primordial Black Holes from XGIS-THESEUS}

\author{Diptimoy Ghosh}
\email{diptimoy.ghosh@iiserpune.ac.in}

\author{Divya Sachdeva}
\email{divya.sachdeva@students.iiserpune.ac.in}

\author{Praniti Singh}
\email{praniti.singh@students.iiserpune.ac.in}

\affiliation{Department of Physics, Indian Institute of Science Education and Research Pune, India}

\begin{abstract}
Current observations allow Primordial Black Holes (PBHs) in asteroid mass range $10^{17}-10^{22}$ g to constitute the entire dark matter (DM) energy density (barring a small mass range constrained by 21\,cm observations). In this work, we explore the possibility of probing PBH with masses $10^{17}-10^{19}\,{\rm g}$ via upcoming X and Gamma  Imaging  Spectrometer (XGIS) telescope array on-board the Transient  High-Energy Sky and Early Universe Surveyor (THESEUS) mission. While our projected limits are comparable with those proposed in the literature for $10^{16}\,{\rm g}\,<\,M_{\mathrm{PBH}}\,<\,10^{17}\,{\rm g}$, we show that the XGIS-THESEUS mission can potentially provide the strongest bound for $10^{17} \mathrm{~g}<M_{\mathrm{PBH}} \lesssim 3\times 10^{18} \mathrm{~g}$ for non-rotating PBHs. The bounds become more stringent by nearly an order of magnitude for maximally rotating PBHs in the mass range $5\times10^{15}\,{\rm g}\,<\,M_{\rm PBH}\,\lesssim\,10^{19}\,{\rm g}$.
\end{abstract}
\maketitle

\section{Introduction}\label{section1}
There are a large variety of astrophysical as well as cosmological observations in support of the Dark Matter (DM) hypothesis. In the absence of any DM candidate within the SM, a variety of models have been proposed to explain this additional matter content, with masses as small as $10^{-22}$ eV (lower bound comes from studies of dwarf galaxy morphology \cite{Chen_2017}). On the high mass end, PBHs are one of the popular candidates of DM that could have formed as a result of large density perturbations in the early universe~\cite{10.1093/mnras/152.1.75,10.1093/mnras/168.2.399,Chapline:1975ojl,Carr:1975qj}. The recent detection of BH merger through direct detection of gravitational waves by LIGO-Virgo~\cite{Abbott_2016,Abbott_2019} have given a new direction to search for PBH as DM~\cite{Bird:2016dcv,Clesse_2017,Sasaki_2016}. 
\\

Regardless of how PBHs are produced, they produce radiation through Hawking evaporation. The energy spectrum of this radiation is almost thermal. An uncharged rotating BH of mass $M_{\mathrm{BH}}$ and angular momentum $J_{\mathrm{BH}}$ radiates with temperature~\cite{PhysRevD.13.198,PhysRevD.14.3260,Hawking:1975vcx,PhysRevD.41.3052}:

\begin{equation}\label{equation}
    T_{\mathrm{BH}}=\frac{1}{4 \pi G_{N} M_{\mathrm{BH}}} \frac{\sqrt{1-a_{*}^{2}}}{1+\sqrt{1-a_{*}^{2}}}
\end{equation}

$G_{N}$ denotes Newton's Gravitational constant and $a_{*}= J_{\mathrm{BH}}/G_{N}^2 M_{\mathrm{BH}}$  is the dimensionless spin parameter of BH. The energy spectrum of the emitted particles peaks around $T_{\mathrm{BH}}$ ($E \sim  5.77\,T_{\mathrm{BH}}$ for photons~\cite{MacGibbon:2007yq,PhysRevD.41.3052})  and then becomes exponentially suppressed for energies greater than $T_{\mathrm{BH}}$ and falls off as a power law for $E\ll T_{\mathrm{BH}}$. Such evaporation causes mass loss on timescales given by:
\begin{equation}
\tau(M) \sim \frac{G^{2} M^{3}}{\hbar c^{4}} \sim \pi\times10^{17}\left(\frac{M_{\mathrm{BH}}}{10^{15} \text{ g.}}\right)^{3} \mathrm{s}
\end{equation}
Consequently, PBHs with mass greater than $10^{15}$ g have lifetime longer than age of the Universe and can contribute to the DM density. 
\\
 \begin{figure}[t]
\includegraphics[width =8.7cm,height=6cm]{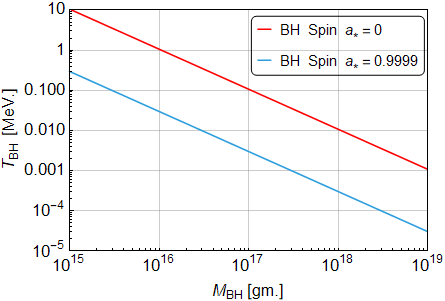}
\caption{\label{fig:temp}BH temperature $T_{\mathrm{BH}}$ as function of its mass $M_{\mathrm{BH}}$ for non-rotating ($a_{*}=0$) and maximally rotating ($a_{*}=0.9999$) case.}
\end{figure} 

Based on the Hawking radiation, there exist cornucopia of constraints from non-observation of emission products from PBH of mass region $\sim 5\times 10^{14}-2\times10^{17}$ g which are presently evaporating. Previous works have constrained this mass range via absence of any excess in electron-positron pairs~\cite{Siegert:2021upf,Dasgupta:2019cae,Laha:2019ssq,DeRocco_2019,Boudaud_2019,Keith:2021guq}, non-observation of neutrino flux from currents detectors~\cite{Dasgupta:2019cae} as well as the upcoming ones~\cite{PhysRevD.81.104019,deromeri2021signatures,calabrese2021primordial,Wang_2021}, the $\gamma$-rays observations~\cite{Ballesteros:2019exr,Carr_2016,Arbey_2020,Laha_2020,Siegert:2021upf}, X-rays observations~\cite{Ballesteros:2019exr,Iguaz_2021}, observation of radio-signals~\cite{Dutta_2021,Chan_2020}, 21-cm observations~\cite{Cang:2021owu,mittal2021constraining,Natwariya:2021xki} and heating of interstellar medium~\cite{Laha_2021,Kim_2021} (for in-depth discussion on similar such studies of PBH, refer to Ref.~\cite{Carr_2020,Carr:2020gox,Green_2021,Carr:2021bzv}). Due to strong constraints on DM fraction, $f_{\rm PBH}$, going as small as $10^{-8}$, this mass range cannot contribute to the total DM density. 

Nonetheless, there exist no constraints in the asteroid mass range $\sim2\times 10^{17}-10^{22}$ g of PBHs as they radiate photons and neutrinos below the threshold sensitivity of the present telescopes and detectors. Therefore, this mass range can entirely correspond to DM~\cite{Smyth_2020,Katz:2018zrn,Montero_Camacho_2019,Niikura:2017zjd,Croon:2020ouk}. Recently, various studies~\cite{Ray:2021mxu,Coogan:2020tuf} have proposed the use of upcoming telescopes/missions like AMEGO, e-ASTROGRAM~\cite{e-ASTROGAM:2017pxr} having larger effective area and lower energy reach, to probe a part of this parameter space. For ex. AMEGO~\cite{AMEGO:2019gny} has been expected to probe mass range upto $\sim 2 \times 10^{18}$ g. giving the most stringent constraints by far. In this work, we study the possibility of probing the DM fraction of non-spinning and spinning PBHs by considering their X-ray and $\gamma$-ray emission, and their detection via X and Gamma Imaging Spectrometer(XGIS)~\cite{Labanti:2021gji} instrument on-board The Transient High-Energy Sky and Early Universe Surveyor\footnote{\url{https://www.isdc.unige.ch/theseus/}} (THESEUS~\cite{THESEUSConsortium:2021yvf}) mission. 
\\

This paper is structured as follows. We begin by reviewing Hawking radiation of a BH in Section \ref{section2} followed by flux calculation from PBH in Section \ref{section3}. Section \ref{section4} describes formalism adopted for our analysis followed by bounds obtained. We conclude in Section \ref{section5}.

\section{Emission rate from BH}\label{section2}
The number ($i$) of particles of spin $s$ and rest mass $m_{0}$ evaporating from BH of mass $M_{\mathrm{BH}}$ and having dimensionless spin parameter $a_{*}$ in energy interval $E$ and $E+dE$ and time interval $dt$ is (in natural units $\hbar=c=k_{B}=1$):
\begin{equation}{\label{spectra}}
    \frac{d^{2} N_{i}}{d E d t}=\sum_{\text {dof }} \frac{1}{2 \pi} \frac{\Gamma_{i}^{lm}\left(E, M_{\mathrm{BH}}, a_{*}, m_{0}, s\right)}{e^{E^{\prime} / T}-(-1)^{2 s}}
\end{equation}
where $E^{\prime}\equiv E-m\Omega$ is the energy of the particle that takes into account the horizon rotation
with angular velocity $\Omega$. The summation is over all degrees of freedom (dof) of particle: color ($i$),  helicity, angular momentum $l$ and its projection $m \in [-l,+l]$. The graybody factor $\Gamma_{i}^{lm}$ encodes the probability that a particle $i$ generated at BH horizon escapes its gravitational well and reaches space infinity \cite{Arbey:2019jmj}. Moreover, $\Gamma_{i}^{lm}$ accounts for the deviation of BH evaporation spectrum from ideal black-body spectrum. In the high energy limit, i.e. $G_{N}M_{\mathrm{BH}}E\gg1$, graybody factor becomes independent of particle's spin $s$ but in the opposite limit, it strongly depends on spin $s$ and is a complicated function of $E$ and $M_{\mathrm{BH}}$. Moreover, it attains the value $\Gamma_{i}^{lm}=27G_{N}^{2}M_{\mathrm{BH}}^{2}E^{2}$ through geometric optics approximation in the high energy limit \cite{PhysRevD.13.198,PhysRevD.16.2402}.
\\

In our analysis, we use public code \verb+BlackHawk v2.0+\footnote{\url{https://blackhawk.hepforge.org/}} \cite{Arbey_2019} for computation of primary (given by eq.\ref{spectra}) and secondary photon spectra. For secondary spectra, the code provides different hadronization schemes given by public codes \verb+Hazma+~\cite{Coogan:2019qpu}, \verb+HERWIG+~\cite{Bahr:2008pv} and \verb+PYTHIA+~\cite{Sjostrand:2006za}. \verb+Hazma+ is accurate for low energy photon spectrum (below QCD confinement scale $\Lambda_{\mathrm{QCD}}$), where pions are fundamental degrees of freedom. On the other hand, for energy range above $\Lambda_{\mathrm{QCD}}$, \verb+PYTHIA+ and \verb+HERWIG+ is suited for computing secondary photon spectra which considers SM particles as elementary degrees of freedom~\cite{Arbey:2021mbl}. Thus, we use \verb+Hazma+ and \verb+PYTHIA+ below and above $\Lambda_{\mathrm{QCD}}\sim200$ MeV respectively.
\\

 \begin{figure}[t]
\includegraphics[width =8.7cm,height=6cm]{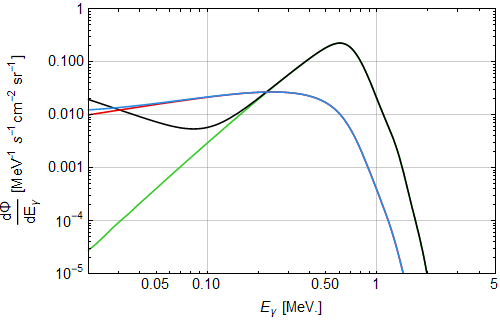}
\caption{\label{fig:2}The Galactic and extra-Galactic photon contribution from non-rotating PBH of mass $10^{17}$ g due to Hawking evaporation assuming they make up the entirety of DM i.e. $f_{\mathrm{PBH}}=1$. The Black (green) and Blue (red) curves correspond to primary (secondary) contribution of  Galactic and extra-Galactic respectively for NFW density profile and region of interest $|l|\leq 5^{\circ}$ and $|b|\leq 5^{\circ}$.}
\end{figure} 
We highlight that \verb+PYTHIA+ gives unreliable results for $E_{\gamma}\leq 5$ GeV since it extrapolates the values from its domain of validity ($\sim$ 5 GeV-100 TeV) \cite{Arbey:2021mbl,Coogan:2020tuf}. Consequently, the extra-Galactic secondary flux contribution would be a source of uncertainty in our calculations, especially for $M_{\rm PBH}\,>\,10^{16}\,{\rm g}$, as dictated by Fig.~\ref{fig:2}.

\section{Flux from PBH}\label{section3}
Assuming monochromatic mass-distribution of PBHs, the differential photon flux per unit solid angle from galactic PBHs having DM fraction $f_{\mathrm{PBH}}$ and mass $M_{\mathrm{PBH}}$ is given by:

\begin{equation}
   \left. \frac{d \Phi_{\mathrm{gal}}}{d E_{\gamma}} \right|_\mathrm{mono,PBH}=\frac{1}{4 \pi} \frac{f_{\mathrm{PBH}}}{M_{\mathrm{PBH}}} \frac{\mathrm{d}^{2} N_{\gamma}}{dE_{\gamma} dt} \int_{\operatorname{LOS}} \rho_{\mathrm{gal}}[r(s,l,b)] ds,
\label{eq:gal}
\end{equation}
where the line of sight integral is given by $J_{\mathrm{D}}$ factor expressed as:
\begin{equation}
    J_{\mathrm{D}} \equiv \frac{1}{\Delta \Omega}\int_{\operatorname{\Delta \Omega}}d\Omega \int_{0}^{s_{\max }} \rho_{\mathrm{gal}}[r(s, l, b)] d s,
\end{equation}
where $\Delta \Omega$ is solid angle of target under observation, $r=\sqrt{r_{\odot}^{2}+s^{2}-2r_{\odot}s\cos(b)\cos(l)}$ being Galacto-centric radius computed as a function of Galactic co-ordinates $(s,l,b)$. In the above expressions, $s$ is the distance from observer, $r_{\odot}=8.3$ kpc is the Galacto-centric distance of the Sun, $\rho_{\mathrm{gal}}(r)$ is the DM profile of Milky Way (MW), $l$ and $b$ are Galactic longitude and latitude respectively. The upper limit of LOS,  $s_{\mathrm{max}}$ is a function of maximum size of MW halo and the direction of the target under observation given as
 \begin{equation}
     s_{\max }=r_{\odot} \cos (b) \cos (l)+\sqrt{r_{\max }^{2}-r_{\odot}^{2}\left(1-\cos ^{2}(b) \cos ^{2}(l)\right)},
 \end{equation}
where $r_{\mathrm{max}}=200$ kpc. In our analysis, we take  Navarro-Frank-White (NFW) profile~\cite{Navarro:1996gj} for Galactic DM density $\rho_{\mathrm{gal}}$ with scale radius, $r_{s}=17$ kpc for Milky Way (MW) halo~\cite{McMillan:2011wd}.

\begin{figure}[h]
\includegraphics[width =8.5cm,height=6cm]{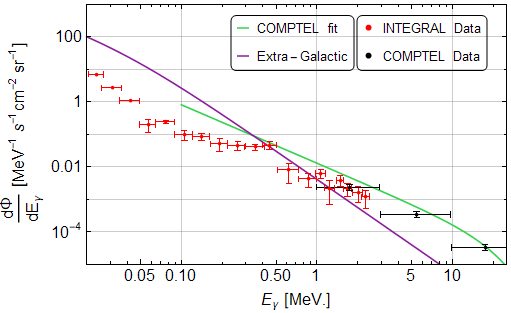}
\caption{\label{fig:bkgd}Galactic and extra-Galactic astrophysical background flux considered in our analysis. Note that INTEGRAL (COMPTEL) data marked in red (black) corresponds to  the region $|l|\,\leq\,30^{\circ}$ and $|b|\,\leq\,15^{\circ}$ whereas the other contribution depicted here correspond to ROI $|l|\,\leq\,5^{\circ},|b|\,\leq\,5^{\circ}$.  }
\end{figure} 

Assuming isotropic distribution of DM at sufficiently large scales, there should be an isotropic photon flux contribution from PBH at all epochs since transparency.  Thus, extra-Galactic differential flux per unit solid angle for monochromatic PBH mass distribution~\cite{Ray:2021mxu} is given by :
 
 \begin{equation}
     \left. \frac{d \Phi_{\mathrm{eg}}}{d E_{\gamma}}\right|_{\mathrm{mono,PBH}}=\frac{1}{4 \pi}\frac{f_{\mathrm{PBH}} \rho_{\mathrm{DM}}}{M_{\mathrm{PBH}}}\left.\int_{0}^{z_{\mathrm{max}}} \frac{d z}{H(z)} \frac{d^{2} N}{d E d t}\right|_{E \rightarrow[1+z] E},
\label{eq:eg}
 \end{equation}
where $\rho_{\mathrm{DM}}$ is the local DM density and $H(z)=H_{0} \sqrt{\Omega_{\Lambda}+\Omega_{m}(1+z)^{3}+\Omega_{r}(1+z)^{4}}$ is Hubble expansion rate at redshift ($z$) and $H_{0}$ is Hubble expansion rate at present epoch. For $z$ integral, the upper limit,  $z_{\mathrm{max}}\approx1100$, which is the redshift for last scattering of CMB photons~\cite{Arbey:2019vqx}. The
current dark-energy, matter, and radiation densities of
the Universe are denoted by $\Omega_{\Lambda},\Omega_{m}$ and $\Omega_{r}$ respectively. We use numerical values of cosmological parameters from latest Planck 2018 results~\cite{Planck:2018vyg}.

\section{Methods \& Results}
\label{section4}
In what follows we examine the sensitivity of the forth-coming telescope array, XGIS, hosted by THESEUS mission to constrain the DM fraction of PBH in mass range $10^{15}-10^{18}$ g. 

\begin{table}[t]
\caption{\label{tab:table1}%
Instrumental details for XGIS detector used in our analysis
}
\begin{ruledtabular}
\begin{tabular}{ccc}
\textrm{Energy Range [MeV.]}&\textrm{Energy Resolution }&
$\Delta E/E$\\
\colrule
 0.02-0.15&20\% FWHM @ 6 keV&$\sim 8.5\%$\\
 0.15-5&6\% FWHM @ 500 keV&$\sim 2.5\%$\\
\end{tabular}
\end{ruledtabular}
\end{table}

\begin{figure}[htb]
\includegraphics[width =8.4cm,height=6cm]{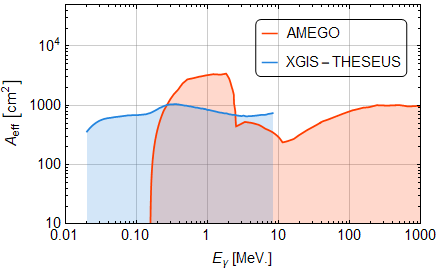}
\caption{\label{fig:effarea}The effective area ($A_{\mathrm{eff}}(E_{\gamma})$), as a function of energy,
of proposed X-ray/gamma-ray telescopes. We have taken simulated $A_{\mathrm{eff}}(E_{\gamma})$ function for XGIS-THESEUS at off-axis angle $0^{\circ}$~\cite{Campana:2021ith}. For AMEGO, the effective area is taken from Ref.~\cite{Coogan:2020tuf}.}
\end{figure} 

\begin{figure*}[htb]
\includegraphics[width =16cm,height=8cm]{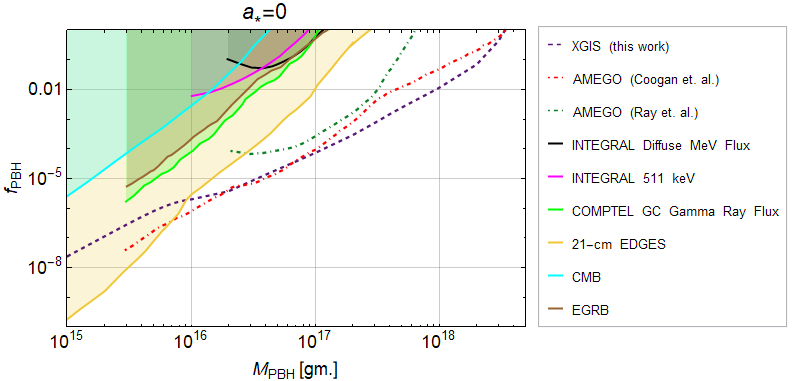}
\caption{\label{fig:Constraints}Projected upper limits (dashed purple) from XGIS telescope array on the DM fraction $f_{\mathrm{PBH}}$, assuming DM is entirely composed of non-rotating PBHs with monochromatic mass distribution. Existing constraints derived from COMPTEL observations (green)~\cite{Coogan:2020tuf} and INTEGRAL observations of Galactic Centre MeV flux (black)~\cite{Laha_2020}, EDGES 21-cm observation (yellow)~\cite{mittal2021constraining}, CMB anisotropy (cyan)~\cite{Acharya_2020}, the extragalatic
gamma-ray background (brown)~\cite{PhysRevD.81.104019} and study of 511 keV line at Galactic Centre (magenta)~\cite{Laha:2019ssq} are shown for comparison. We have also shown projected bounds for AMEGO telescope derived in Ref.~\cite{Ray:2021mxu} (dot-dashed green) and Ref.~\cite{Coogan:2020tuf} (dot-dashed red).}
\end{figure*} 

\begin{figure*}[t]
\includegraphics[width =16cm,height=8cm]{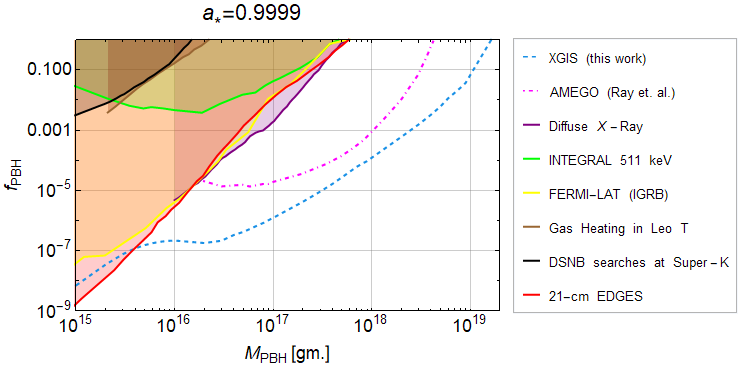}
\caption{\label{fig:RotConstraints}Projected upper limits (dashed blue) from XGIS telescope array on the DM fraction $f_{\mathrm{PBH}}$, assuming DM is entirely composed of maximally-rotating PBHs with monochromatic mass distribution. Existing constraints derived from study on isotropic X-ray and soft gamma-ray background (purple)~\cite{Iguaz_2021}, 511 keV line at Galactic Centre (green) \cite{Dasgupta:2019cae}, FERMI-LAT isotropic gamma-ray background (yellow)~\cite{Arbey:2019vqx}, gas heating in LeO T dwarf galaxy (brown) \cite{Laha_2021}, diffuse supernova neutrino background (DSNB) searches at Super-Kamiokande (black) \cite{Dasgupta:2019cae} and 21-cm EDGES observation (red) \cite{Cang:2021owu} are shown for comparison. We have also shown projected bounds for AMEGO telescope (magenta dot-dashed) \cite{Ray:2021mxu}}
\end{figure*} 

{\it Astrophysical Backgrounds}: We consider various observations of extra-galactic as well as galactic X-ray or gamma-ray backgrounds in the energy range [20 {\rm keV}, 5 {\rm MeV}] where the lower end of the energy range is  determined by the fact that effective area for XGIS~\cite{Labanti:2021gji} is given only for $E_\gamma \gtrsim 20\,{\rm keV}$ (see Fig.~\ref{fig:effarea}). Note that our bounds would become more stringent when considering the entire range (i.e [2 {\rm keV}, 10 {\rm MeV}]). In our analysis, we have taken target under observation to be Galactic Centre given by: $|l|\leq 5^{\circ}$ and $|b|\leq5^{\circ}$.

For extra-Galactic astrophysical background in the energy range [20 {\rm keV}, 5 {\rm MeV}], we have considered double power-law model proposed in Ref.~\cite{Ballesteros:2019exr}. This model fits the cosmic X-ray background spectrum and is based on observations from SMM~\cite{doi:10.1063/1.53933}, Nagoya balloon~\cite{Fukada1975}, HEAO–1 and HEAO-A4~\cite{Kinzer_1997,Gruber_1999}:
\begin{equation}
   \left. \frac{d \Phi_{\mathrm{eg}}}{d E_{\gamma}} \right|_{\mathrm{bkg}}=\frac{A_{\mathrm{eg}}}{\left(E_{\gamma}/ E_{0}\right)^{n_{1}}+\left(E_{\gamma}/ E_{0}\right)^{n_{2}}}
\end{equation}
where $n_{1}=1.4199$ and $n_{1}=2.8956$, $E_{0}$= 35.6966 keV and $A_{\mathrm{eg}}$= 64.2 $\mathrm{MeV}^{-1} \mathrm{~cm}^{-2}\mathrm{~s}^{-1}\mathrm{sr}^{-1}$. 

The galactic background in the energy range $0.1$ to $5\,{\rm MeV}$ for the region of interest i.e, $|l|\leq 5^{\circ}$ and $|b|\leq 5^{\circ}$, can be parameterized as~\cite{Bartels:2017dpb}:
\begin{equation}
     \left. \frac{d \Phi_{\mathrm{gal}}}{d E_{\gamma}} \right|_{\mathrm{bkg}}=\displaystyle 0.013\left(\frac{E_\gamma}{1 \mathrm{MeV}}\right)^{-1.8}  e^{-\left(\frac{E_\gamma}{20 \mathrm{MeV}}\right)^{2}}, 
\end{equation}
given in units of $\mathrm{~cm}^{-2} \mathrm{~s}^{-1} \mathrm{sr}^{-1} \mathrm{MeV}^{-1}$. This background model fit is in accordance with the data obtained by COMPTEL~\cite{Strong:1998ck,Strong:2011pa,1994AA}. For energy less than a $0.1\,{\rm MeV}$, we use data obtained with the SPI instrument onboard INTEGRAL~\cite{Bouchet:2011fn} taken in the region $|l|\,\leq\,30^{\circ}$ and $|b|\,\leq\,15^{\circ}$. 

Fig.~\ref{fig:bkgd} shows the Galactic and extra-Galactic astrophysical backgrounds used in this analysis. Note that it is evident from Fig.~\ref{fig:bkgd}, the extra-galactic background is an order of magnitude larger than the galactic background for energies less than a $0.1\,{\rm MeV}$. This leads to a difference of 8.5\% in the results obtained for no. of photons observed by the XGIS detector with and without INTEGRAL data for energy range $20\,{\rm keV}-0.1\,{\rm MeV}$, in addition to extra-galactic background. 

{\it Detector Characteristics and analysis}: 
The main characteristic of XGIS detector are taken from Ref.~\cite{Labanti:2021gji} and are given in Table~\ref{tab:table1}. The values of energy efficiency $\epsilon(E_\gamma)$ (or $\Delta E_\gamma/E_\gamma$) are given as $\sim0.424$ times the FWHM values.  Note that we extend the values of $\epsilon$ given at particular energy values (at 6 keV and 500 keV) in  Ref.~\cite{Labanti:2021gji} in the respective energy ranges. Furthermore, the energy and angular resolution are not crucial for our analysis.

In Fig.~\ref{fig:effarea}, we have shown effective area ($A_{\rm eff}$) of detector XGIS and we compare this with that of AMEGO. Note that in comparison to various telescopes considered in Ref.~\cite{Coogan:2020tuf}, XGIS detector sensitivity extends even below soft gamma ray regime (till $\sim$ 20 keV). For now, it has been proposed that the two camera of XGIS will be positioned at offset angle $\pm20^{\circ}$ with respect to the observation axis of the satellite~\cite{Labanti:2021gji}, reducing $A_{\mathrm{eff}}$ which in turn makes PBH constraints weaker by a factor of $\sim 1.1$.

Equipped with various parameters of XGIS telescope array, next, we define the number of photons observed by the detector in energy range $[E_{\mathrm{min}},E_{\mathrm{max}}]$ as 
\begin{equation}
    N_{\gamma}=\displaystyle T_{\mathrm{obs}} \int\displaylimits_{E_{\mathrm{min}}}\displaylimits^{E_{\max }} d E_{\gamma} A_{\mathrm{eff}}\left(E_{\gamma}\right) \int d E_{\gamma}^{\prime} R_{\epsilon}\left(E_{\gamma} \mid E_{\gamma}^{\prime}\right) \frac{\mathrm{d} \Phi}{\mathrm{d} E_{\gamma}^{\prime}}
    \label{eq:nop}
\end{equation}
where $T_{\mathrm{obs}}$ is observation time and $ R_{\epsilon}\left(E_{\gamma} \mid E_{\gamma}^{\prime}\right)$ is spectral resolution function. The spectral resolution function accounts for the probability that true photon energy $E_{\gamma}^{\prime}$ is assigned to value $E_{\gamma}$ after reconstruction by the detector which can be approximated as Gaussian distribution~\cite{Bringmann:2008kj}:
\begin{equation}
R_{\epsilon}\left(E_{\gamma} \mid E_{\gamma}^{\prime}\right) \equiv \frac{1}{\sqrt{2 \pi} \epsilon\left(E_{\gamma}^{\prime}\right) E_{\gamma}^{\prime}} \exp \left[-\frac{\left(E_{\gamma}-E_{\gamma}^{\prime}\right)^{2}}{2\left(\epsilon\left(E_{\gamma}^{\prime}\right) E_{\gamma}^{\prime}\right)^{2}}\right]
\end{equation}

We use Eq.~\ref{eq:gal} and Eq.~\ref{eq:eg} in Eq.~\ref{eq:nop} to obtain the detected no. of photons of signal, considering contribution of primary and secondary evaporation spectra to the flux of Galactic and extra-Galactic emission from PBH. Similarly, we estimate the no. of photons corresponding to total astrophysical background, $N_{\gamma}|_{\rm{bkg}}$. The projected bounds are, then, derived by demanding signal-to-noise ratio (SNR) over the time of observation, $T_{\mathrm{obs}}=10^{8}$ s to be greater than five~\cite{Coogan:2020tuf,Coogan:2019qpu}:
\begin{equation}
 \frac{\left. N_{\gamma}  \right|_{\mathrm{PBH}}}{\sqrt{\left. N_{\gamma}  \right|_{\mathrm{bkg}}}} \geq 5.
\end{equation}
Note that this requirement corresponds to the discovery of any signal from PBH as the sole candidate of DM. We present the projected upper limits on DM fraction of non-rotating and maximally rotating PBHs, $f_{\rm PBH}$, that can be obtained from future XGIS-THESEUS  observations in Fig.~\ref{fig:Constraints} and Fig.~\ref{fig:RotConstraints} respectively. We have also shown projected limits obtained from AMEGO and other existing constraints for the comparison.

We highlight that the shape of the curve obtained for XGIS is identical to that of GECCO in Ref.~\cite{Coogan:2020tuf} since both are sensitive to lower energy gamma-rays and have comparable effective area $A_{\mathrm{eff}}$. We obtain stronger constraints for $M_{\rm PBH}\,>\,10^{16}$ g  compared to the lower mass region since the lower masses have peak energy outside the energy range for XGIS (0.02-5 MeV). For maximally rotating case, the limits are more stringent and extend to larger masses of PBH. We found that XGIS can exclude PBH as exclusive DM component upto mass $\sim 2\times10^{18}$ g for non-rotating PBH and $\sim 1.5\times10^{19}$ g for maximally rotating.

\section{Summary and Outlook}\label{section5}
PBHs with masses less than $10^{17}$ g are severely constrained via non-observation of Hawking evaporation products, most stringest constraints by far is given by 21 cm observation in Ref.~\cite{mittal2021constraining}. On the other hand, PBHs in asteroid mass range, as of now, can entirely consist of DM as they radiate significantly in low energy range (below MeV scale) where most present telescopes lack the required sensitivity. In this work, we showed how future telescope like XGIS with their larger effective area and low energy threshold can probe some part of this parameter space. We included contribution of both primary and secondary spectra to the flux of Galactic and extra-Galactic emissions from PBHs in our analysis. Assuming that the astrophysical background measured by XGIS will be similar to that of observed by INTEGRAL and COMPTEL, we derived the projected bounds by demanding SNR over the time of observation greater than five for non-rotating and rotating PBHs in Fig.~\ref{fig:Constraints} and Fig.~\ref{fig:RotConstraints} respectively. We found that the mission can give stronger bound than current constraints for $10^{17}\,{\rm g}\,<\,M_{\rm PBH}\,\lesssim\,3\times10^{18}\,{\rm g}$ for non-rotating PBHs. For maximally rotating PBHs, we obtained limits an order of magnitude larger than the non-rotating case extending to larger range of PBH masses. It should be noted that more information on the effective area for lower energies can lead to stronger limits. Thus, our projections are conservative. We conclude that XGIS-THESEUS has the potential to either probe or impose stronger constraints than the existing ones, in some part of the parameter space of asteroid mass range PBHs. 

\section{ACKNOWLEDGEMENTS}
DG and DS acknowledge support through the Ramanujan Fellowship and MATRICS Grant of the Department of Science and Technology, Government of India. We also acknowledge International Centre for Theoretical Sciences (ICTS) for organizing online program - Less Travelled Path of Dark Matter: Axions and Primordial Black Holes\footnote{\url{https://www.icts.res.in/program/LTPDM2020}} (code: ICTS/LTPDM2020/11), whose talks and discussions played a vital role in this work.
\bibliographystyle{apsrev4-1.bst}
\bibliography{reference}

\end{document}